\documentclass[aps,prl,twocolumn,amsmath,amssymb,floatfix,superscriptaddress,showpacs]{revtex4}
\usepackage{graphicx}

\begin{document}


\title{Artificial kagom\'e arrays of nanomagnets: a frozen dipolar spin ice}



\author{N.~Rougemaille}
\affiliation{Institut N\'eel, CNRS-UJF, BP 166, 38042 Grenoble Cedex 9, France}
\author{F.~Montaigne}
\affiliation{Institut Jean Lamour, CNRS-Nancy Universit\'e, BP 239, F-54506 Vandoeuvre, France}
\author{B.~Canals}
\affiliation{Institut N\'eel, CNRS-UJF, BP 166, 38042 Grenoble Cedex 9, France}
\author{A.~Duluard}
\affiliation{Institut Jean Lamour, CNRS-Nancy Universit\'e, BP 239, F-54506 Vandoeuvre, France}
\author{D.~Lacour}
\affiliation{Institut Jean Lamour, CNRS-Nancy Universit\'e, BP 239, F-54506 Vandoeuvre, France}
\author{M.~Hehn}
\affiliation{Institut Jean Lamour, CNRS-Nancy Universit\'e, BP 239, F-54506 Vandoeuvre, France}
\author{R.~Belkhou}
\affiliation{Synchrotron SOLEIL, L'Orme des Merisiers Saint-Aubin, 91192 Gif-sur-Yvette, France}
\affiliation{ELETTRA, Sincrotrone Trieste, I-34012 Basovizza, Trieste, Italy}
\author{O.~Fruchart}
\affiliation{Institut N\'eel, CNRS-UJF, BP 166, 38042 Grenoble Cedex 9, France}
\author{S.~El~Moussaoui}
\affiliation{Synchrotron SOLEIL, L'Orme des Merisiers Saint-Aubin, 91192 Gif-sur-Yvette, France}
\affiliation{ELETTRA, Sincrotrone Trieste, I-34012 Basovizza, Trieste, Italy}
\author{A.~Bendounan}
\affiliation{Synchrotron SOLEIL, L'Orme des Merisiers Saint-Aubin, 91192 Gif-sur-Yvette, France}
\affiliation{ELETTRA, Sincrotrone Trieste, I-34012 Basovizza, Trieste, Italy}
\author{F.~Maccherozzi}
\affiliation{Synchrotron SOLEIL, L'Orme des Merisiers Saint-Aubin, 91192 Gif-sur-Yvette, France}
\affiliation{ELETTRA, Sincrotrone Trieste, I-34012 Basovizza, Trieste, Italy}


\date{\today}

\begin{abstract}
Magnetic frustration effects in artificial kagom\'e arrays of nanomagnets are investigated using X-PEEM microscopy and monte carlo simulations. Spin configurations of demagnetized networks reveal unambiguous signatures of long range, dipolar interaction between the nanomagnets. As soon as the system enters the spin ice manifold, the kagom\'e dipolar spin ice model captures the observed physics, while the short range kagom\'e spin ice model fails.
\end{abstract}

\maketitle


Complex architectures of nanostructures are nowadays routinely elaborated using bottom-up and/or micro-fabrication processes. This technological capability allows scientists to engineer materials with properties that do not exist in nature, such as negative refractive index \cite{Shelby01} or tunable photonic bandgap \cite{Busch99}, but also to elaborate model systems to explore fundamental issues in condensed matter physics. Frustrated spin models are one example for which theoretical predictions can be tested experimentally \cite{Gingras09}. In these models, spins are placed on an infinite network of a given geometry, and interact through a short or long range coupling of ferromagnetic or antiferromagnetic nature. Under certain conditions, magnetic frustration effects occur, giving rise to an infinite degeneracy of the ground state, i.e. a finite entropy at zero temperature. Several experimental studies on two-dimensional, artificial arrays of ferromagnetic nanomagnets have been reported recently \cite{Wang06,Tanaka06,Qi08, Zabel08,Nolting08}, bringing together the communities of magnetic frustration, statistical physics and nanomagnetism. Properties of these arrays seem to be well described by short range spin ice (SI) models: they show disorder, satisfy ice-like rules, and correlation coefficients fit with predictions \cite{Qi08}.

In the kagom\'e SI model, spins interact through first neighbor exchange coupling only. Experimentally however, nanomagnets are coupled via the magnetostatic interaction, which extends beyond nearest neighbors. While the main interest for frustrated compounds arises from the massive degeneracy of their ground state, this degeneracy is fully lifted when long range, dipolar interactions are included in the model \cite{finally_ordered_Moller09,finally_ordered_Tchernyshyov09}. Understanding whether or not these long range interactions influence the local spin configuration in artificial arrays of nanomagnets is thus essential, especially because these networks are often considered as a playground to study magnetic frustration effects on a mesoscopic scale. Combining experimental and theoretical investigations, we show that the kagom\'e dipolar spin ice (DSI) model captures the physics we observe, where the kagom\'e SI model fails.

\begin{figure}
\includegraphics[width=8.5cm]{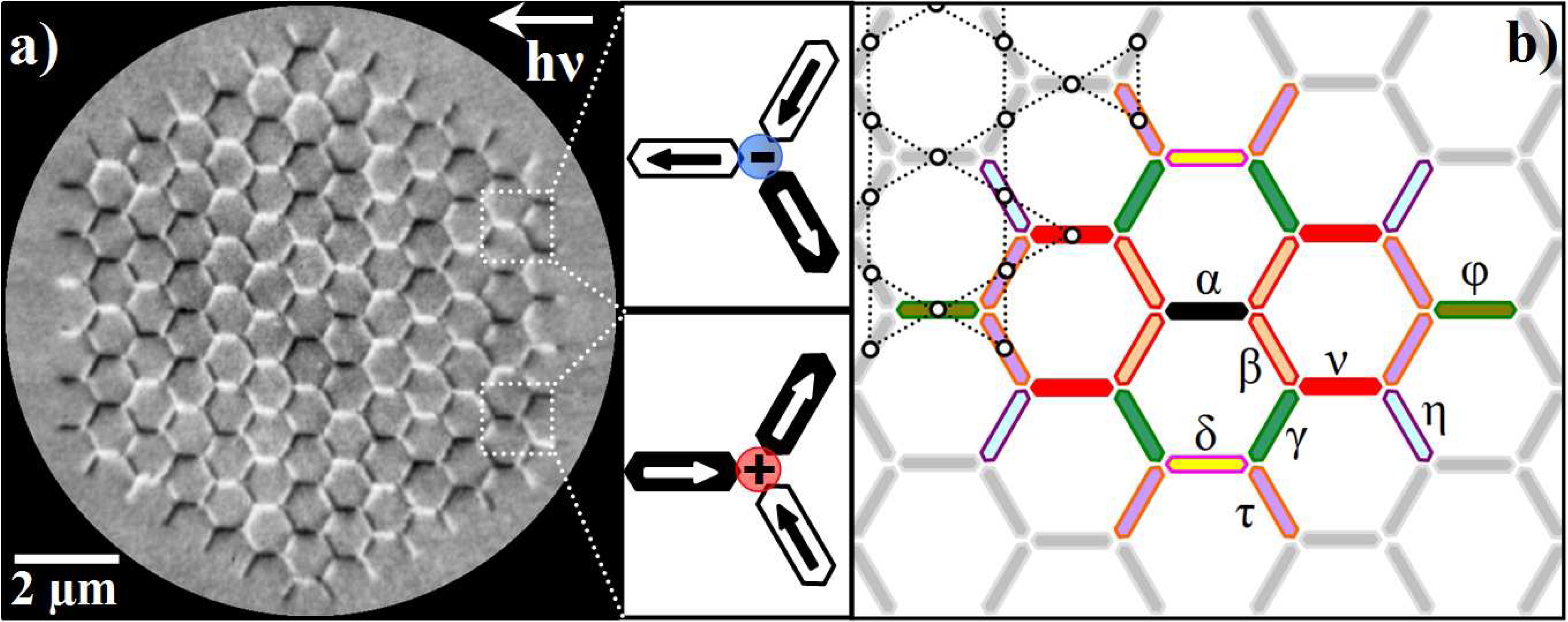}
\caption{\label{fig1}
(Color online) (a) 15 $\mu$m X-PEEM magnetic image of a demagnetized network. Direction of magnetization is deduced from the black and white contrast, as illustrated for two vertices that satisfy the ice rule, i.e. the local 1in/2out (magnetic charge Q=$-$1) or 2in/1out (magnetic charge Q=$+$1) spin configuration. (b) Sketch of the array indicating the relative indices used for the 7 nearest neighbors. The partly superimposed dots highlight the nodes of the kagom\'e lattice. 
}
\end{figure}

Honeycomb arrays of nanomagnets have been fabricated from a Si//Ta(5nm)/Co(10nm)/Ru(2nm) thin film. An aluminium mask, defined by e-beam lithography and lift-off, has been used to protect the structures during ion beam etching. The Al mask has been subsequently removed by chemical etching. The nanomagnets have a typical size of $470\times70\times10$ nm$^3$, ensuring a high uniaxial shape anisotropy, so that they can be considered, locally, as Ising pseudo spins. The distance between nearest nodes in the network is set to 500 nm (Fig. 1a). The Co nanomagnets are thus physically disconnected, and are coupled through the magnetostatic interaction. The networks of nodes have kagom\'e topology (Fig. 1b) and are composed of 342 Co nanomagnets. The magnetic configuration of each element is resolved using the X-ray PhotoEmission Electron Microscope (X-PEEM, SOLEIL instrument) of the Nanospectroscopy beamline at the ELETTRA synchrotron facility (Trieste, Italy). Magnetic images are acquired at the Co $L_3$ edge at remanence, after demagnetizing the networks ex situ. The importance of the demagnetizing procedure to approach the ground state spin configurations has been highlighted in previous works \cite{Wang07, Ke08}. To demagnetize our networks, the sample is rotated with a period of about 20 ms in a damped, in-plane sinusoidal magnetic field (1 s period, 0.11 Oe amplitude variation between two periods). From the X-PEEM images, the magnetic state of each individual nanomagnet is measured (Fig. 1a). Data presented here have been obtained from 9 nominally identical networks, which have been demagnetized 6 times, out of which 34 magnetic configurations have been determined.

 \begin{table}
 \caption{\label{table}
Theoretical (SI model) and experimental spin-spin correlators, number of $i j$ interactions in our networks, and coupling coefficients in arbitrary unit deduced from the point-dipole approximation and from micromagnetic calculations.
 }
\begin{ruledtabular}
 \begin{tabular}{|c|c|c|c|c|c|c|c|}
    & $\alpha \beta$	& $\alpha \gamma$ & $\alpha \nu$ & $\alpha \delta$ & $\alpha \tau$ & $\alpha \eta$ & $\alpha \phi$ \\
\hline
$C_{i j}$ (theo.)    & 0.167 & -0.062 & 0.101 & -0.075 & 0.012 & 0.019 & 0.023 \\
\hline
$C_{i j}$ (exp.)    & 0.164 & -0.056 & 0.151 & -0.063 & 0.013 & 0.056 & 0.021 \\
\hline
$N_{i j}$    & 648 & 612 & 612 & 303 & 1152 & 576 & 273 \\
\hline
$J_{i j}$ (dip.)    & 1 & -0.137 & 0.045 & -0.036 & 0.014 & 0.037 & 0.014 \\
\hline
$J_{i j}$ ($\mu$mag.)    & 4.905 & -0.130 & 0.044 & -0.030 & 0.013 & 0.040 & 0.015 \\
 \end{tabular}
 \end{ruledtabular}
 \end{table}

The kagom\'e ice rule, i.e. the preferential selection of 1in/2out or 2in/1out low-energy spin configurations (see Fig. 1a), is globally well obeyed: over all magnetic configurations, less than 0.6\% of the vertices have a 3in or 3out high-energy spin configuration. These "forbidden" vertices are likely due to structural defects in the networks and often appear at the same locations. Residual magnetization after demagnetization is low, of the order of a few percents, consistent with refs. \onlinecite{Wang07} and \onlinecite{Ke08}. From the analysis of our magnetic images, we deduce the spin-spin correlation coefficients $C_{ij} = \langle \vec{S}_i.\vec{S}_j \rangle$ between normalized $\vec{S}_i $ and $\vec{S}_j$ spins (indices are relative, identical to those used in previous literature \cite{Wills02,Qi08}, and defined in Fig. 1b). The experimental values, averaged over the 34 demagnetized networks, are reported in Table \ref{table} for the 7 nearest neighbors. These values are compared to the coefficients expected from the SI model, described by the Hamiltonian $H_{\textrm{SI}}=-J_{\alpha \beta} \, \sum_{\langle i j \rangle}\vec{S}_i . \vec{S}_j $ that couples nearest neighbors only, through the $J_{\alpha \beta}$ coefficient \cite{Wills02,Note_Benj}. Although differences are visible, as previously observed \cite{Qi08}, experimental results and theoretical predictions from the SI model agree relatively well at first sight.

Considering the nanomagnets as Ising spins \cite{Note_François}, we can derive from the $C_{ij}$ values the network energy $E=-(1/N) \, \sum_{i \neq j}^{} N_{i j} J_{i j} C_{i j}$. In this expression, $N$ is the total number of spins, $N_{i j}$ and $J_{i j}$ being respectively the number of interactions and coupling coefficients of type $i j$ (Fig. 1b). Within the point-dipole approximation, the coupling coefficients $J_{i j}$ can be calculated: for two magnetic moments $\mu \, \vec{S}_i$ and $\mu \, \vec{S}_j$ located at $\vec{r}_i$ and $\vec{r}_j$:
$$
J_{i j} = - D \, ( \, \frac{1}{|\vec{r}_{ij}|^{3}}
- 3 \frac{( \vec{S}_i . \vec{r}_{ij} ) ( \vec{S}_j . \vec{r}_{ij} )}{|\vec{r}_{ij}|^{5} \, \vec{S}_i . \vec{S}_j} \, ),
$$
with $\vec{r}_{ij} = \vec{r}_i - \vec{r}_j$, and $D = ( \mu_0 / 4 \pi) \mu^2$ the strengh of the dipolar coupling. The corresponding $J_{i j}$ values are reported in Table \ref{table} and compared to the values obtained using micromagnetic simulations \cite{Oommf}. Good agreement is found between the two approaches, except for first neighbor interactions, for which the minimal distance between the nanomagnets is small compared to their extension. 

\begin{figure}
\includegraphics[width=8.7cm]{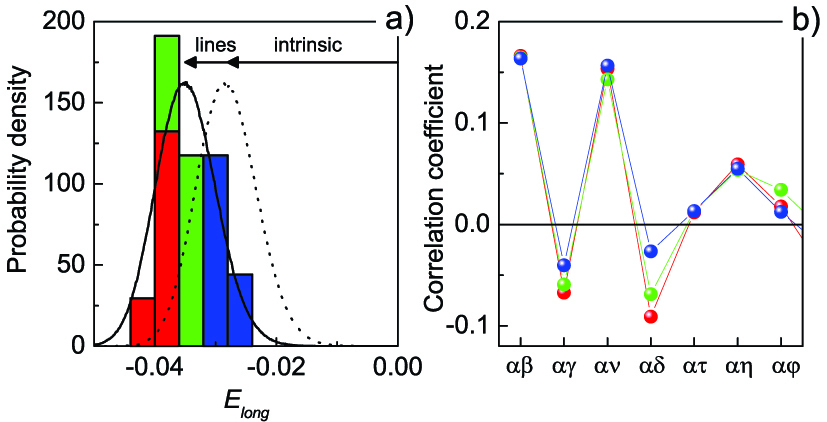}
\caption{\label{fig2}
(Color online) (a) Experimental histogram deduced from the 34 demagnetized networks. Color code is used to sort the configurations in 3 subsets. Curves show the normalized $E_{\textrm{long}}$ distributions from the SI model (dotted line) and after replacing the expected $C_{\alpha \nu}$ and $C_{\alpha \eta}$ coefficients by their measured values (continuous line). (b) Experimental correlation coefficients averaged within each energy subset.
}
\end{figure}

The system energy can be split in two terms: $E = E_{\textrm{short}} + E_{\textrm{long}}$ where $E_{\textrm{short}}$ is the energy related to the interaction between first neighbors only ($E_{\textrm{short}} = -N_{\alpha \beta} J_{\alpha \beta} C_{\alpha \beta} /N$), and $E_{\textrm{long}}$ the energy related to the interactions beyond the nearest neighbors. We will show that, once the system enters the spin ice manifold, $C_{\alpha \beta}$ is configuration-independent, i.e $E_{\textrm{short}}$ is constant. In the following, we thus focus our discussion on the $E_{\textrm{long}}$ term. The presence of a few "forbidden" vertices in our networks  is then not critical as $E_{\textrm{long}}$ excludes energy contributions from nearest neighbors. The experimental $E_{\textrm{long}}$ distribution is shown in Fig. 2a, and compared to the distribution expected from the SI model (dotted line). Interestingly, this distribution is shifted towards negative energy. To understand where energy is gained in our networks, we sort the 34 measured configurations in 3 subsets corresponding to the lowest (red), intermediate (green) and highest (blue) energy configurations (Fig. 2a), and we average the $C_{ij}$ coefficients within these 3 subsets (Fig. 2b). Two main features are then observed: i) independently of the configuration energy, the $C_{\alpha \nu}$ and $C_{\alpha \eta}$ coefficients  have higher values than expected (also see Table 1), and ii) the values of the $C_{\alpha \gamma}$ and $C_{\alpha \delta}$ coefficients strongly vary with energy, contrary to what is observed for the 5 other measured coefficients.

As depicted in Fig. 1b, the $C_{\alpha \nu}$ and $C_{\alpha \eta}$ coefficients reflect correlations along a line in the network. We might think that these higher correlation coefficients originate from the demagnetization procedure that may favor ferromagnetically coupled nanomagnets along these lines. However, although the system has threefold symmetry, the three main directions of lines are not equivalent, and we observe anisotropy in the $C_{\alpha \nu}$ and $C_{\alpha \eta}$ values (one direction is different from the two others). We thus attribute the high $C_{\alpha \nu}$ and $C_{\alpha \eta}$ values to variations of coupling coefficients between first neighbors, presumably due to anisotropy in the fabrication process. Replacing the theoretical correlation coefficients along a line by the experimental ones, keeping all the other coefficients unchanged, leads to a larger shift of the distribution towards negative energies (continuous line in Fig. 2a). This new energy shift reproduces well our data. Thus, the total energy shift we observe experimentally has two main origins: an intrinsic shift associated with the ice model itself ($E_{\textrm{long}}$ is lower in the SI manifold than in the paramagnetic state in which all spin configurations are equiprobable \cite{note3}), and an extrinsic shift due to the lithography process. Both phenomena are related to first neighbor interactions, and do not involve long range coupling.

In contrast with the other correlation coefficients, the values of $C_{\alpha \gamma}$ and $C_{\alpha \delta}$ strongly vary with $E_{\textrm{long}}$ (see Figs. 2b). For several configurations, $C_{\alpha \gamma}$ and $C_{\alpha \delta}$ reach absolute values that can be surprisingly large compared to the values expected from the kagom\'e SI model. Magnetostatic coupling between the ($\alpha, \gamma$) and ($\alpha, \delta$) spins favors antiparallel alignment of the corresponding nanomagnets, and thus negative values of their spin-spin correlators. Intuitively, this could explain why the measured $C_{\alpha \gamma}$ and $C_{\alpha \delta}$ values are not in quantitative agreement with the values expected from the kagom\'e SI model.

\begin{figure}
\includegraphics[width=8.2cm]{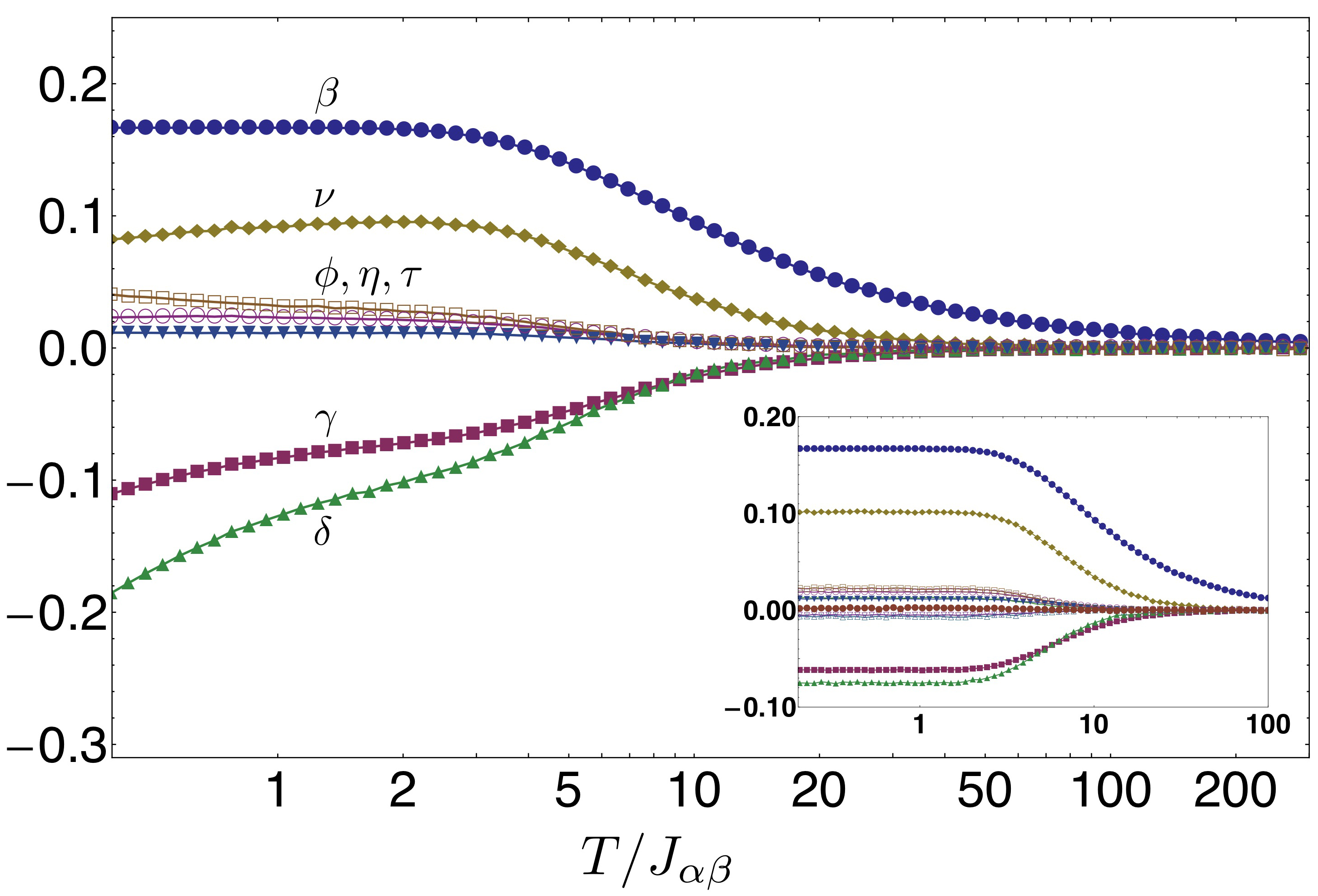}
\caption{\label{fig3}
(Color online) Correlation coefficients $C_{\alpha i}$ derived from the DSI model as a function of temperature $T$. Greek index $i=\beta, \gamma, \ldots$ identifies each correlator. Inset : temperature dependance of correlators within the SI model.}
\end{figure}

To test whether this interpretation better describes our findings, monte carlo simulations of the kagom\'e DSI model have been performed, using the following Hamiltonian: 
$H_{\textrm{DSI}}= - J_1 \sum_{\langle i  j \rangle} \vec{S}_i . \vec{S}_j - \sum_{i \neq \j} J_{i j} \, \vec{S}_i . \vec{S}_j$, where $J_1$ is an additional isotropic nearest neighbor coupling. This $J_1$ term can be tuned so that the effective nearest neighbor interaction $J_1 + J_{\alpha \beta} = 5 \, J_{\alpha \beta}$, to account for the spatial extension of the nanomagnets (see Table 1). The simulations were done for $L \times L \times 3$ sites lattices ($L$ ranging from 11 to 144) with periodic boundary conditions, using a single spin flip algorithm and a simulated annealing procedure from $T/J_{\alpha \beta} = 100$ to $T/J_{\alpha \beta} = 0.1$. In these simulations, 10$^5$ modified monte carlo steps (mmcs) are used for thermalization, where 1 mmcs corresponds to a set of local updates sufficiently long to achieve stochastic decorrelation, itself determined using the on the fly calculated spin-spin autocorrelation time. Measurements follow the thermalization process and are computed with 10$^5$ mmcs. Results of the simulation are shown in Fig. 3 where spin-spin correlators are plotted as a function of temperature. In the high temperature regime, the system is in a paramagnetic state. As temperature is lowered, the system enters the short range kagom\'e spin ice manifold ($C_{\alpha \beta} = 1/6$), and the 7 measured correlation coefficients nearly reach the values expected from the SI model. We emphasize that in absence of long range interaction, these values would then remain temperature-independent (see inset of Fig. 3). The situation is different when dipolar interactions are included in the calculation. At very low temperatures, a transition to an ordered phase takes place \cite{finally_ordered_Moller09,finally_ordered_Tchernyshyov09}: all correlation coefficients vary abruptly (not shown here), and strongly deviate from the SI values. In the intermediate regime, i.e when $2 \le T/J_{\alpha \beta} \le 5$ (see Fig. \ref{fig3}), mean values of the correlation functions do not allow easy discrimination between the SI and DSI models. However, two correlation coefficients slowly decrease when lowering the temperature (see the two lowest curves). These two coefficients, $C_{\alpha \gamma}$ and $C_{\alpha \delta}$, are precisely those for which we observe a change experimentally, when $E_{\textrm{long}}$ decreases. This strikingly contrasts with predictions from the SI model and fits well our experimental data.

An alternative description of these spin models is useful. Identifying each spin $S_i$ to a magnetic dipole allows the mapping of any spin configuration on the kagom\'e lattice onto a classical magnetic charge configuration on the honeycomb lattice (Fig. 1a). The high temperature paramagnetic phase of the SI and DSI models then translates into a paramagnetic charge state. When temperature is lowered, both models enter the spin ice manifold: the kagom\'e ice rules are satisfied, and charge values are constrained to $\pm 1$ (instead of $\pm 1$ and $\pm 3$ at high temperature), therefore corresponding to an emergent $\mathbb{Z}_2$ field. In the SI model, this Ising field remains weakly correlated and temperature independent. The DSI model however behaves differently: at lower temperatures, it leaves a first spin ice manifold (SI-I) and enters a more constrained phase (SI-II), in which spins still fluctuate but where charges crystallize in an antiferromagnetic pattern \cite{finally_ordered_Tchernyshyov09}. During the phase transition of the hidden $\mathbb{Z}_2$ field, the stochastic dynamics is strongly modified, with exponentially suppressed single spin flips, leaving only loop-like moves to operate efficiently in the SI-II manifold. We have thus investigated the temperature dependance of the short range magnetic charge correlation $\langle Q_i.Q_{i+1} \rangle$, and compare monte carlo predictions from the SI and DSI models (Fig. 4a) to our measurements (Fig. 4b). In the highest temperature regime, where all charge states are accessible (Q=$\pm$1 and Q=$\pm$3), the unconstrained paramagnetic charge phase is associated with an asymptotic correlator $\langle Q_i.Q_{i+1} \rangle$\,=\,-1 (Note that a Q=$\pm$3 charge state corresponds to a monopole defect, see for example refs. \onlinecite{Ladak10} and \onlinecite{Mengotti10}). When the temperature is lowered, the SI model enters the spin ice manifold, with constrained $Q=\pm 1$ charges and a temperature independant short range charge correlation $\langle Q_i.Q_{i+1} \rangle$=-1/9. In the DSI model however, the charge correlator first increases to reach a maximum value around -0.15, and then steadily decreases with temperature. Note that there exists a temperature range within the spin ice manifold, in which the mean values of the correlator, as well as their associated standard deviations, are mutually exclusive for the two models (Fig. 4a). Analysis of our X-PEEM images reveals strong deviations from the -1/9 value of the SI model (Fig. 4b), clearly outside standard deviations of the correlator distributions, and agrees well with predictions from the DSI model. We emphasize that in all 34 networks, the spin configurations are charge neutral, which rules out a global extrinsic breaking of the $\mathbb{Z}_2$ symmetry. Therefore, we attribute the deviations of the short range charge correlations to the proximity of the SI-II regime, in which the $\mathbb{Z}_2$ and translation symmetries are broken. Unambiguously, our results reveal that dipolar interactions are at work in our arrays.

\begin{figure}
\includegraphics[width=8.4cm]{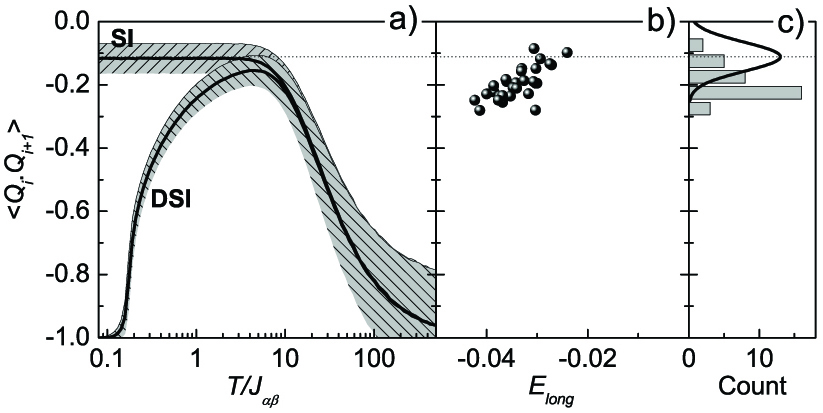}
\caption{\label{fig4}
(a) Theoretical short range magnetic charge correlators $\langle Q_i.Q_{i+1} \rangle$ (bold line) and their standard deviations (hatched regions) calculated from the SI and DSI models as a function of temperature. (b) Experimental $\langle Q_i.Q_{i+1} \rangle$ values measured for the 34 demagnetized networks as a function of $E_{\textrm{long}}$. (c) Histogram of the experimental charge correlators compared to the nearly gaussian distribution computed in the SI model, within the spin ice manifold.}
\end{figure}

In summary, numerical computation and magnetic imaging show that the kagom\'e SI model fails to interpret our experiments, while the kagom\'e DSI model captures the physics we observe. The fact that artificial arrays of nanomagnets may look like short range kagom\'e spin ice systems at first sight is fortuitous, and related to the critical slowing down at low temperature of the single spin flip dynamics. Therefore, while artificial kagom\'e arrays of nanomagnets should not exhibit finite entropy at zero temperature \cite{finally_ordered_Moller09, finally_ordered_Tchernyshyov09}, they are most likely described by a frozen dipolar spin ice regime with an emergent symmetry breaking of the hidden $\mathbb{Z}_2$ field associated with the classical magnetic charges.


\begin{thebibliography}{2}
\bibitem{Note_Benj}
Note that the values of our spin-spin correlators are slightly different from those reported in Ref. \cite{Wills02}.
\bibitem{Note_François}
This assumption is far from trivial for a micromagnetic object. We checked for our structures that the internal magnetic configuration is essentially independent on the external magnetic configuration. The energy of the system can then be expressed as a sum of pair interactions. Material parameters are: exchange = $3\times10^{-11}$ J/m, saturation magnetization = $1.4\times10^3$ kA/m, and magnetocrystalline anisotropy = 0. Cell size is $1\times1\times10$  nm$^3$. Elements have a size of $470\times70\times10$ nm$^3$. A detailed analysis will be presented in a forthcoming publication.
\end{thebibliography}
\end{document}